\documentclass[pra,twocolumn,showpacs,superscriptaddress]{revtex4}
\usepackage{graphicx}
\begin{document} 
 
\title{Groverian Entanglement Measure of Pure Quantum States
with Arbitrary Partitions}
 
\author{Yishai Shimoni and Ofer Biham} 
\affiliation{Racah Institute of Physics, 
The Hebrew University, Jerusalem 91904, Israel} 
 
\begin{abstract} 

The Groverian entanglement measure of pure quantum states
of $n$ qubits is generalized to the case in which 
the qubits are divided into any $m \le n$ parties and the 
entanglement between these parties is evaluated.
To demonstrate this measure we apply it to general states of 
three qubits and to symmetric states with any number of 
qubits such as the Greenberg-Horne-Zeiliner state and the W state.

\end{abstract} 
 
 
\maketitle 
 
\section{Introduction} 
\label{sec:introduction}

The potential speedup offered by quantum computers is exemplified
by Shor's factoring algorithm
\cite{Shor1994}, Grover's search algorithm 
\cite{Grover1996,Grover1997a},
and algorithms for
quantum simulation
\cite{Nielsen2000}.
Although the origin of 
this speed-up is not fully understood, 
there are indications that quantum entanglement 
plays a crucial role
\cite{Jozsa2003,Vidal2003}. 
In particular,
it was shown that quantum algorithms that do not create entanglement 
can be simulated efficiently on a classical computer
\cite{Aharonov1996}.
It is therefore of interest to quantify the entanglement 
produced by quantum algorithms and examine its correlation
with their efficiency.  
This requires to develop entanglement measures for the quantum states
of multiple qubits that appear in quantum algorithms.

The special case of bipartite entanglement 
has been studied extensively in recent years.
It was established as a resource for 
quantum teleportation procedures.
The entanglement of pure bipartite states can be evaluated by the
von Neumann entropy of the reduced density matrix,
traced over one of the parties.
For mixed bipartite states, several measures were proposed,
namely entanglement of formation and entanglement of distillation
\cite{Bennett1996a,Bennett1996b}.
In particular, for 
states of two qubits an exact formula for the
entanglement of formation was obtained
\cite{Hill1997,Wootters1998}.
Bipartite pure states of more than two qubits were also studied.
It was shown that generic quantum states can be reconstructed
from a fraction of the reduced density matrices, obtained by
tracing over some of the qubits
\cite{Linden2002a,Linden2002b}.

The more general case of multipartite entanglement is
not as well understood.
Recent work based on axiomatic considerations has provided a set
of properties that entanglement measures should satisfy
\cite{Vedral1997,Vedral1998,Vidal2000,Horodecki2000}.
These properties include the requirement that any entanglement
measure should vanish for product (or separable) states, 
it should be invariant under local unitary operations
and should not increase
as a result of any sequence of local operations 
complemented by only classical 
communication between the parties.
Quantities that satisfy these properties are called entanglement
monotones.
These properties, 
that should be satisfied for bipartite as well as multipartite entanglement,
provide useful guidelines in the search for
entanglement measures for multipartite quantum states.
One class of entanglement measures, 
based on metric properties of the Hilbert 
space was proposed and shown to satisfy these requirements
\cite{Vedral1997,Vedral1997a,Vedral1998}.
Another class of measures, based on polynomial invariants
has been studied in the context of multipartite entanglement
\cite{Barnum2001,Leifer2004}.
However, the connection between such measures and the efficiency
of quantum algorithms remains unclear.

The Groverian measure of entanglement for 
pure quantum states of
multiple qubits
provides an operational interpretation 
in terms of the success probability of certain quantum
algorithms
\cite{Biham2002}.
More precisely,
the Groverian measure of a state 
$| \psi \rangle$
it is related to the success 
probability of Grover's search
algotirhm when this state 
is used as the initial state. 
A pre-precessing stage is allowed 
in which an arbitrary local unitary operator is applied to each
qubit.
These operators are optimized in order to
obtain the maximal success probability 
of the algorithm,
$P_{\rm max}$.
The Groverian measure is given by 
$G(|\psi\rangle) = \sqrt{1 - P_{\rm max}}$
\cite{Biham2002}.
For a state 
$| \psi \rangle$
of $n$ qubits,
the entanglement evaluated by this measure is, in fact,
the entanglement between $n$ parties,
where each of them holds a single qubit.
The Groverian measure has been used in order to characterize
quantum states of high symmetry such as the 
Greenberg-Horne-Zeilinger (GHZ)  
and the W states
\cite{Shimoni2004}.
It has also been used to evaluate the entanglement produced by 
quantum algorithms such as Grover's algorithm
\cite{Shimoni2004} and
Shor's algorithm
\cite{Shimoni2005}.
The Groverian measure was also generalized to the case of
mixed states
\cite{Shapira2006}.

Consider a quantum state
$| \psi \rangle$
of $n$ qubits.
These qubits can be partitioned into any
$m \le n$
parties, each holds one or more qubits.
In this paper we present a 
generalized Groverian measure 
which quantifies the 
parties for any desired partition.
This is done
by allowing 
any unitary operators within each partition.
This essentially changes the meaning of locality 
to encompass the whole party,
enabling a more complete characterization of 
quantum states of multiple qubits.

The paper is organized as follows.
In Sec. 
\ref{sec:algorithm} 
we briefly describe Grover's search algotirhm.
In Sec.  
\ref{sec:measure} 
we  review the Groverian entanglement measure. 
In Sec. 
\ref{sec:generalized} 
we present the generalized Groverian measure that applies for
any desired partition of the quantum state.
In Sec.  
\ref{sec:numerical}
we present an efficient numerical procedure for the
calculation of the generalized Groverian measure. 
We use this measure in Sec.
\ref{sec:results}
to characterize certain pure quantum states
of high symmetry.
A brief discussion is presented in Sec. 
\ref{sec:discussion}.
The results are summarized in
Sec. 
\ref{sec:summary}.

\section{Grover's Search Algorithm}
\label{sec:algorithm}

Grover's algorithm 
performs a search for a marked element $m$ in 
a search space $D$ containing $N$ elements.
We assume, for convenience, 
that $N = 2^n$, where $n$ is an integer.
This way, the elements of $D$ can be represented by 
an $n$-qubit register
$| x \rangle = | x_1,x_2,\dots,x_n \rangle$, 
with the computational basis states 
$| i \rangle$, $i=0,\dots,N-1$.
The meaning of marking the element $m$, 
is that there is a function
$f: D \rightarrow \{0,1\}$, 
such that $f=1$ for the marked elements, and $f=0$ for the rest. 
To solve this search 
problem on a classical computer one needs to evaluate $f$ for each 
element, one by one, until the marked state is found.  Thus, on average, 
$N/2$ evaluations of $f$ are required and $N$ in the worst case.
On a quantum computer, where $f$ can be evaluated \emph{coherently},
a sequence of unitary operations,
called Grover's algorithm
and denoted by
$U_G$, 
can locate a marked element using only $O(\sqrt{N})$ coherent 
queries of $f$ \cite{Grover1996,Grover1997a}.
The algorithm is based on a unitary operator, called a quantum oracle,
with the ability to recognize the marked states.
Starting with the equal superposition state, 

\begin{equation}
|\eta\rangle=\sum_{i=0}^{N-1}|i\rangle,
\end{equation}

\noindent 
and applying the operator $U_G$
one obtains the state

\begin{equation}
U_G |\eta\rangle = |m\rangle + O({1}/{N}),
\label{eq:<mU}
\end{equation}

\noindent
which is then measured.
The success probability of the algorithm 
is almost unity.
The adjoint equation takes the form
$\langle\eta| = \langle m|U_G + O({1}/{N})$.
If an arbitrary pure state, $|\psi\rangle$,
is used as the initial state 
instead of the state $| \eta \rangle$,
the success probability is reduced
to

\begin{equation}
P_s = 
|\langle m|U_G|\psi\rangle|^2 + O({1}/{N}).
\label{eq:Psmpsi}
\end{equation}

\noindent
Using Eq.~(\ref{eq:<mU}) we obtain

\begin{equation}
P_s=|\langle\eta|\psi\rangle|^2 + O({1}/{N}),
\label{eq:etaU}
\end{equation}

\noindent
namely, the success probability is determined by the overlap
between 
$| \psi \rangle$
and the equal superposition
$| \eta \rangle$
\cite{Biham2002,Biham2003}.

\section{The Groverian Entanglement Measure}
\label{sec:measure}

\begin{figure}
\includegraphics[width=\columnwidth]{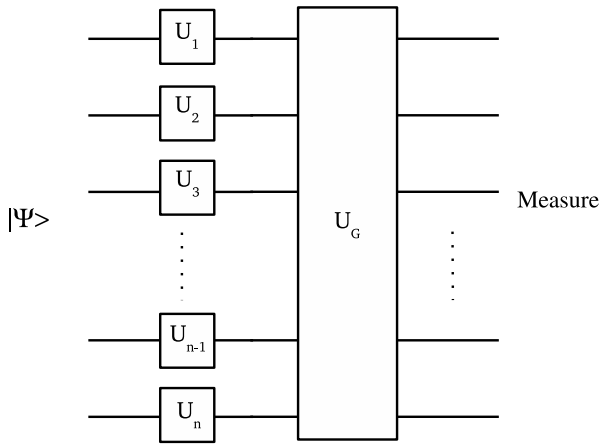}
\caption{
The quantum circuit that exemplifies the operational 
meaning of the Groverian entanglement measure $G(\psi)$.
A pure state 
$| \psi \rangle$
of $n$ qubits is inserted as the input state.
In the pre-processing stage, a local unitary operator
is applied to each qubit before the resulting state
is fed into Grover's algorithm.
The local unitary operators
$U_i$, $i=1,\dots,n$
are optimized in order to maximize the success 
probability of the search algorithm for the given
initial state
$| \psi \rangle$.
}
\label{fig1}
\end{figure}

Consider Grover's search algorithm, in which an arbitrary pure state
$| \psi \rangle$ is used as the initial state.
Before applying the operator $U_G$, there is a pre-processing stage
in which arbitrary local unitary operators
$U_1$, $U_2$, $\dots$, $U_n$
are applied on the $n$ qubits in the register
(Fig. \ref{fig1}).
These operators are chosen such that the success probability
of the algorithm would be maximized.
The maximal success probability is thus given by

\begin{equation}
P_{\rm max} = 
\max_{U_1,U_2,\dots,U_n}
|\langle m|U_G(U_1\otimes\dots\otimes U_n)|\psi\rangle|^2.
\label{eq:Pmax}
\end{equation}

\noindent
Using Eq.~(\ref{eq:<mU}), this can be re-written as

\begin{equation}
P_{\rm max} = 
\max_{U_1,U_2,\dots,U_n}|\langle\eta|U_1\otimes\dots\otimes U_n|\psi\rangle|^2,
\end{equation}

\noindent
or

\begin{equation}
P_{\rm max} = 
\max_{|\phi\rangle \in T}|\langle\phi|\psi\rangle|^2,
\end{equation}

\noindent
where $T$ is the space of all tensor product states of the form

\begin{equation}
|\phi\rangle = |\phi_1\rangle\otimes\dots\otimes|\phi_n\rangle.
\label{eq:T}
\end{equation}

The Groverian measure is given by

\begin{equation}
G(\psi) = 
\sqrt{1 - P_{\rm max}},
\label{eq:G(psi)}
\end{equation}

\noindent
For the case of pure states, for which $G(\psi)$
is defined, it is closely related to an entanglement measure 
introduced in Refs.
\cite{Vedral1997,Vedral1997a,Vedral1998} 
for both pure and mixed states
and was shown to be an entanglement monotone.
This measure can be interpreted as the distance 
between the given state and the nearest separable state.
It is expressed in terms of the fidelity between the
two states.
Based on these results, it was shown 
\cite{Biham2002}
that $G(\psi)$
satisfies
(a) $G(\psi) \geq 0$, with equality only when $|\psi\rangle$
is a product state;
(b) $G(\psi)$ cannot be increased using local operations 
and classical communication
(LOCC). 
Therefore, $G(\psi)$ is an entanglement monotone
for pure states.
A related result was obtained in Ref.
\cite{Miyake2001},
where it was shown that the evolution of the quantum state  
during the iteration of Grover's
algorithm corresponds
to the shortest path in Hilbert space
using a suitable metric.

\section{The Generalized Groverian Measure}
\label{sec:generalized}

\begin{figure}
\includegraphics[width=\columnwidth]{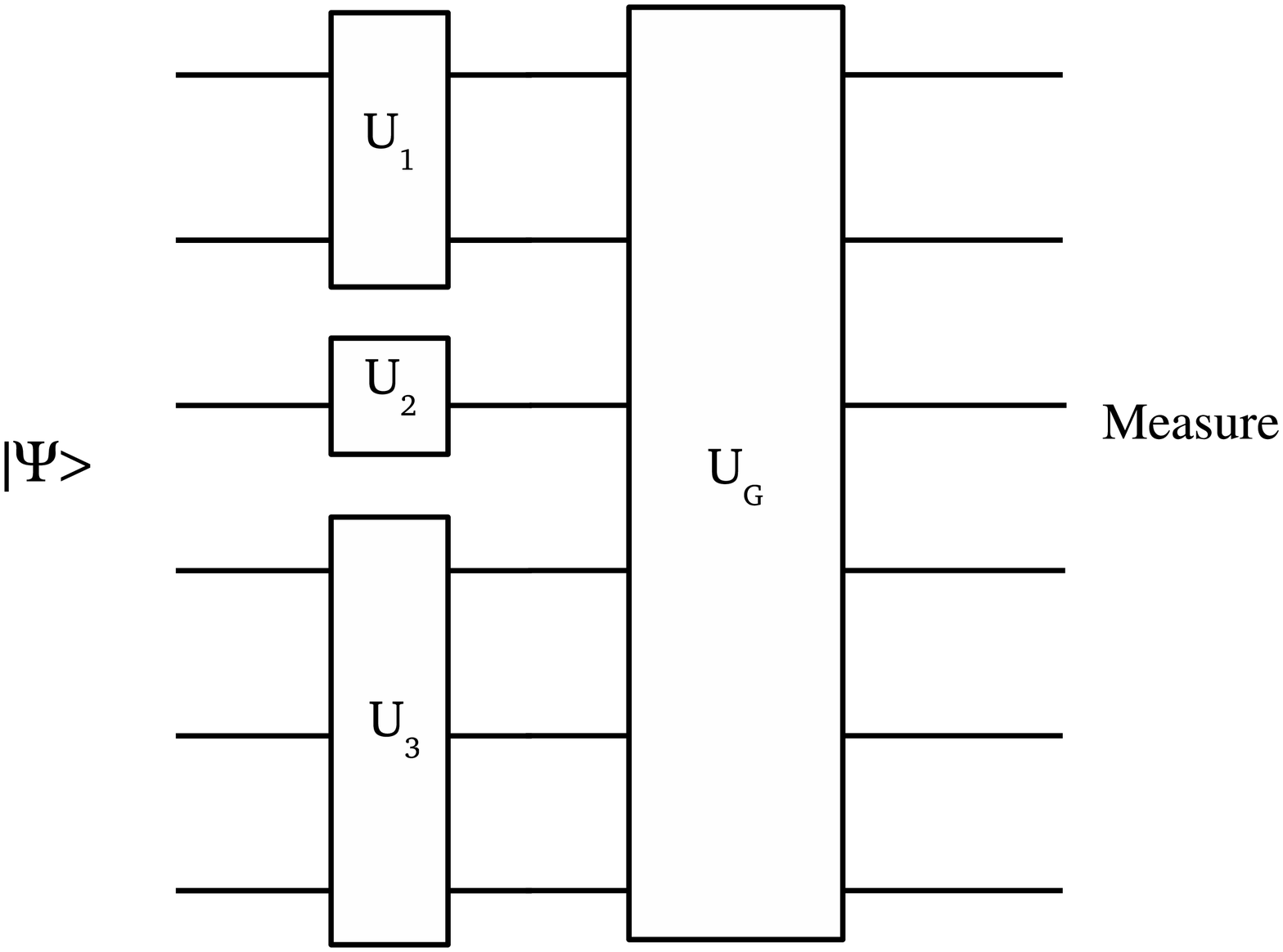}
\caption{
The quantum circuit that exemplifies the operational 
meaning of the generalized Groverian measure $G(\psi)$,
for $n$ qubits divided in a certain way between $m$ parties.
In this example, a pure state 
$| \psi \rangle$
of six qubits,
which is divided between three parties,
is inserted as the input state into Grover's algorithm.
In the pre-processing stage, a local unitary operator
is applied on the qubits held by each party 
before the resulting state
is fed into Grover's algorithm.
The local unitary operators
$U_1, U_2, U_3$
are optimized in order to maximize the success 
probability of the search algorithm,
for the given initial state
$| \psi \rangle$.
}
\label{fig2}
\end{figure}

Consider a quantum state
$|\psi \rangle$ 
of $n$ qubits.
In the original Groverian measure each qubit belongs to a separate
party. The measure quantifies the entanglement between all these 
parties. This is a natural partitioning scheme for states 
created by quantum algorithms. 
The resulting measure
can be considered as an intrinsic property of the state itself.
However, consider a situation in which  
$m \le n$ different parties share the quantum state,
where each party holds one or more qubits.
These parties
wish to cooperate and perform Grover's search 
algorithm on the whole state.
In this situation, in order to maximize the success probability,
the operators $U_i$, $i=1,\dots,m$, 
should no longer be limited to single qubits.
Instead, the operator $U_i$ applies on all the qubits in partition $i$. 
This enables to quantify the inter-party entanglement,
removing the intra-party entanglement. 
The quantum circuit that demonstrates the evaluation of the 
generalized Groverian measure for the state
$|\psi \rangle$ 
with any desired partition
is shown in Fig. 
\ref{fig2}.
The generalized Groverian measure is given 
by Eq.
(\ref{eq:G(psi)})
where
Eq.
(\ref{eq:T})
is replaced by
$|\phi\rangle = |\phi_1\rangle\otimes\dots\otimes|\phi_m\rangle$,
where $| \phi_i \rangle$
is a state of partition $i$.
Clearly, the generalized Groverian measure 
is an entanglement monotone.

\section{Numerical Evaluation of the Generalized Groverian Measure}
\label{sec:numerical}

For a given partition of $m$ parties,
the generalized Groverian measure is expressed in terms 
of the maximal success probability

\begin{equation}
P_{\rm max} = 
\max_{|\phi_i\rangle}|\langle\phi_1|\otimes\dots
\otimes\langle\phi_m|\psi\rangle|^2,
\end{equation}

\noindent
where the maximization is over all possible states 
$| \phi_i \rangle$
of each partition, $i$.
This calls for a convenient parametrization of 
the state of each partition.
Consider a partition $i$ that includes one qubit.
The state of this partition can be expressed by

\begin{equation}
|\phi_i\rangle = 
e^{i \gamma_0} \cos{\theta_0}|0\rangle
+ e^{i \gamma_1} \sin{\theta_0}|1\rangle.
\label{eq:singleq}
\end{equation}

\noindent
In case that the partition includes two qubits, 
its state can be expressed by

\begin{eqnarray}
|\phi_i\rangle &=& 
e^{i \gamma_0} \cos{\theta_0} |0\rangle
+ \sin{\theta_0} [ e^{i \gamma_1} \cos{\theta_1} |1\rangle \nonumber \\
&+& \sin{\theta_1} ( e^{i \gamma_2} \cos{\theta_2} |2\rangle
+ e^{i \gamma_3} \sin{\theta_2} |3\rangle )].
\label{eq:twoqubits}
\end{eqnarray}

\noindent
This parametrization can be generalized to any number
of qubits in partition $i$.
Using this parametrization, 
one can express the overlap function

\begin{equation}
f = \langle\phi_1|\otimes\dots
\otimes\langle\phi_m|\psi\rangle,
\end{equation}

\noindent
in terms of the
$\theta_k$'s and
$\gamma_k$'s of all the partitions.
In fact, $f$ is simply a sum of products of sine, cosine
and exponential functions of the $\theta_k$'s and 
$\gamma_k$'s.
At this point, the steepest descent algorithm 
can be applied to maximize
$|f|$.
However, a more efficient maximization procedure can be
obtained as follows.

For a given partition, one can express $f$ as a function
of $\theta_0$ and $\gamma_0$, fixing all the other parameters
$\theta_k$ and $\gamma_k$
at this and all other partitions,
in the form

\begin{equation}
f=c_0\sin{\theta_0}+e^{i\gamma_0}d_0\cos{\theta_0}.
\end{equation}

\noindent
The values of 
$c_0=|c_0|e^{i\alpha_0}$ 
and 
$d_0=|d_0|e^{i\beta_0}$
depend on all the fixed parameters.
The maximization of $|f|^2$ vs. 
$\theta_0$ and $\gamma_0$ 
leads to

\begin{equation}
|f|^2 \rightarrow |c_0|^2+|d_0|^2.
\end{equation}

\noindent
The values of
$\gamma_0$ 
and
$\theta_0$ 
at which this maximization is obtained are

\begin{eqnarray}
\gamma_0 &\rightarrow& \beta_0 - \alpha_0 \nonumber \\
\cos{\theta_0} &\rightarrow& 
\frac{|c_0|}{\sqrt{|c_0|^2+|d_0|^2}},
\label{eq:params}
\end{eqnarray}

\noindent
where the sign of $\theta_0$ is the same as  
the sign of $|d_0| - |c_0|$. 

Note that the ordering of the states within each partition is
arbitrary. Therefore, 
in order to perform the same procedure for 
$\theta_1$ and $\gamma_1$, 
the parametrization of the two-qubit
partition in Eq.
(\ref{eq:twoqubits})
can be changed to

\begin{eqnarray}
|\phi_i\rangle &=& 
e^{i \gamma_1} \cos{\theta_1} |1\rangle
+ \sin{\theta_1} [ e^{i \gamma_2} \cos{\theta_2} |2\rangle \nonumber \\
&+& \sin{\theta_2} ( e^{i \gamma_3} \cos{\theta_3} |3\rangle
+ e^{i \gamma_0} \sin{\theta_3} |0\rangle )].
\label{eq:twoqubits1}
\end{eqnarray}

In practice, the optimization procedure 
consists of iterations of the following steps:
(a) Randomly choose a basis state 
$| p \rangle$ 
in one of the $m$ partitions;
(b) Reparamietrize the state of the 
chosen partition such that $| p \rangle$ 
will be the left-most state in Eq.
(\ref{eq:twoqubits}); 
(c) Reset $\theta_p$ and $\gamma_p$ in the chosen partition
according to Eq.
(\ref{eq:params}) to maximize $|f|^2$,
while fixing all the other parameters.

\section{Results}
\label{sec:results}

Using the numerical tools described above,
it is possible to evaluate the generalized Groverian entanglement 
of any pure quantum state for any given partition.
Here we demonstrate this approach for pure quantum states of
high symmetry, namely the generalized GHZ state and the W state.

Consider the generalized GHZ state of three qubits

\begin{equation}
| \psi \rangle = a_0 | 000 \rangle + a_1 | 111 \rangle.
\label{eq:GHZ}
\end{equation}

\noindent
The three-party case, in which each party holds one qubit
was considered before
\cite{Shimoni2004}.
It was found that

\begin{equation}
P_{\rm max} = \max(|a_0|^2,|a_1|^2).
\label{eq:GHZpmax}
\end{equation}

\noindent
We will now evaluate the generalized Groverian measure for the
case in which one party holds two qubits and the second
party holds a single qubit.
A general pure state of the first party can be expressed by

\begin{eqnarray}
| \phi_1 \rangle &=& 
  e^{i \gamma_0} \cos{\theta_0} | 00 \rangle
+ e^{i \gamma_1} \sin{\theta_0} \cos{\theta_1} | 01 \rangle \nonumber \\
&+& e^{i \gamma_2} \sin{\theta_0} \sin{\theta_1} \cos{\theta_2} | 10 \rangle 
\nonumber \\
&+& e^{i \gamma_3} \sin{\theta_0} \sin{\theta_1} \sin{\theta_2} | 11 \rangle,
\end{eqnarray}

\noindent
while
a general pure state of the second party is given by

\begin{equation}
| \phi_2 \rangle = 
  e^{i \gamma_4} \cos{\theta_4} | 0 \rangle
+ e^{i \gamma_5} \sin{\theta_4} | 1 \rangle.
\end{equation}

\noindent
The overlap function will take the form

\begin{eqnarray}
f &=&   e^{i \gamma_0} e^{i \gamma_4} \cos{\theta_0} \cos{\theta_4} a_0
\nonumber \\
  &+&   e^{i \gamma_3} e^{i \gamma_5} \sin{\theta_0} \sin{\theta_1} 
      \sin{\theta_2} \sin{\theta_4} a_1. 
\end{eqnarray}

\noindent
The maximization of $|f|^2$ vs. all the 
$\theta_i$'s and $\gamma_i$'s will lead to
Eq. (\ref{eq:GHZpmax}).
This means that for the generalized GHZ state, the generalized Groverian
measure does not depend on the partition. It can be shown that
this result applies to generalized GHZ states with any number of qubits
and any partition.
This can be interpreted as if generalized GHZ states carry only 
bipartite entanglement, in agreement with previous studies
\cite{Dur2000}.

Another family of highly symmetric pure states of multiple qubits
is the class of W states.
The W state of $n$ qubits is given by

\begin{equation}
| \psi \rangle = \frac{1}{\sqrt{n}} \sum_{i=0}^{n-1} | 2^i \rangle,
\end{equation}

\noindent
namely it is the equal superposition of all basis states  
in which one qubit is 1 and all the rest are 0.
This class of states was found to have  
$P_{\rm max} = (1-1/n)^{n-1}$ 
\cite{Shimoni2004}.

We will now extend this analysis to more general partitions of the
$n$-qubit W state.
First, we consider the bipartite case.
In this case, the generalized Groverian entanglement 
is equal to the maximal eigenvalue of the reduced density matrix,
traced over one of the two parties
\cite{Biham2002}.
Consider the simple case in which one party includes a single
qubit, while the other party includes all the other qubits.
In this case we find that $P_{\rm max}=(n-1)/n$.
In the general two-party case, one party includes
$k$ qubits and the other includes $n-k$ qubits.
In this case we find that
$P_{\rm max} = \max(k/n,1-k/n)$.

For more that two parties, the analogy between the generalized
Groverian measure and the largest eigenvalue
of the reduced density matrix does not apply.
Thus, the evaluation of the generalized Groverian measure
can be performed analytically for a few simple cases,
and in general requires the computational procedure described above.

Consider the $n$-qubit W state. 
Here we focus on a simple set of partitions to $m$ parties,
in which $m-1$ parties include one qubit each, and
the last party includes all the remaining qubits.
In Table 
\ref{table1} 
we present 
$P_{\max}$ for W states of $n=2,\dots,7$ qubits
divided between $m=1,\dots,n$ parties.
The results in the first two rows as well as the main diagonal
were obtained analytically as well as by the numerical procedure.
The rest of the results were obtained numerically.
Those results that appear as exact integer fraction were identified
as such based on the numerical results.
In four other cases, we could not identify such exact fractions.

\begin{table}[t]
\caption{The success probability $P_{\max}$ obtained for states of
the W class.
Each column corresponds to the W states with a given number of
qubits, $n = 1,\dots,7$.
Each row corresponds to a given number of partitions,
$m=1,\dots,n$.
Since there can be many ways to partition $n$ qubits
into $m$ parties, we focused on a specific class of partitions
in which $m-1$ parties hold one qubit each and all the remaining
qubits are in one party.
}

\begin{tabular}{|l|ccccccc|}
\hline
Partitions & 1 bit & 2 bits & 3 bits & 4 bits & 5 bits & 6 bits & 7 bits \\
\hline
1 & 1 & 1   & 1         & 1         & 1         & 1         & 1         \\
2 &   & 1/2 & 2/3       & 3/4       & 4/5       & 5/6       & 6/7       \\
3 &   &     & $(2/3)^2$ & 2/4       & 3/5       & 4/6       & 5/7       \\
4 &   &     &           & $(3/4)^3$ & 0.4408    & 3/6       & 4/7       \\
5 &   &     &           &           & $(4/5)^4$ & 0.4198    & 0.4494    \\
6 &   &     &           &           &           & $(5/6)^5$ & 0.4084    \\
7 &   &     &           &           &           &           & $(6/7)^6$ \\
\hline
\end{tabular}
\label{table1}
\end{table}

\section{Discussion}
\label{sec:discussion}

Consider a pure quantum state 
$| \psi \rangle$
of $n$ qubits.
The number of ways 
to divide these $n$ qubits into $m$ parties
is given by the binomial coefficient $C_n^m$.
For each of these partitions, one can evaluate the
generalized Groverian measure 
$G(\psi)$, 
that quantifies the
$m$-partite entanglement between these parties.
In this analysis, locality is defined according to the
partition, so that all the operations that are performed
within a single partition are considered as local.
Using this approach, 
one can identify the partition for which
$G(\psi)$ is maximal among all the partitions that include
$m$ parties,
and denote its value as 
$G_m(\psi)$. 
This quantity satisfies a monotonicity relation of the form
$G_m(\psi) \le G_{m+1}(\psi)$, 
where $G_1(\psi)=0$.
This means that splitting of parties tends to increase this
measure of multipartite
entanglement while merging of parties tends to decrease it.

Furthermore, the interesting question of state ordering 
may be addressed using this measure.
It would be interesting to find pairs of states,
$|\psi_1\rangle$ and $|\psi_2\rangle$, 
such that
$G_{m_1}(\psi_1) < G_{m_1}(\psi_2)$ 
but 
$G_{m_2}(\psi_1) < G_{m_2}(\psi_2)$
for some integers $m_1$ and $m_2$.

\section{Summary}
\label{sec:summary}

In summary, we have presented a generalization of 
the Groverian entanglement measure of multiple quibits
to the case in which the qubits are divided into
any desired partition. The generalized measure 
quantifies the multipartite entanglement between these partitions.
To demonstrate this measure we evaluated it for a variety
of pure quantum states using a combination of analytical and 
numerical methods. In particular, we have studied the
entanglement of highly symmetric states of multiple qubits 
such as the generalized GHZ states and the W states.

\bibliographystyle{apsrev}

\begin{thebibliography}{27}
\expandafter\ifx\csname natexlab\endcsname\relax\def\natexlab#1{#1}\fi
\expandafter\ifx\csname bibnamefont\endcsname\relax
  \def\bibnamefont#1{#1}\fi
\expandafter\ifx\csname bibfnamefont\endcsname\relax
  \def\bibfnamefont#1{#1}\fi
\expandafter\ifx\csname citenamefont\endcsname\relax
  \def\citenamefont#1{#1}\fi
\expandafter\ifx\csname url\endcsname\relax
  \def\url#1{\texttt{#1}}\fi
\expandafter\ifx\csname urlprefix\endcsname\relax\def\urlprefix{URL }\fi
\providecommand{\bibinfo}[2]{#2}
\providecommand{\eprint}[2][]{\url{#2}}

\bibitem[{\citenamefont{{P.W. Shor}}(1994)}]{Shor1994}
\bibinfo{author}{\bibnamefont{{P.W. Shor}}}, in
  \emph{\bibinfo{booktitle}{{Proceedings of the 35th Annual Symposium on the
  Foundations of Computer Science}}}, edited by
  \bibinfo{editor}{\bibnamefont{{S. Goldwasser}}} (\bibinfo{publisher}{{IEEE
  Computer Society}}, \bibinfo{address}{{Los Alamitos, CA}},
  \bibinfo{year}{1994}), p. \bibinfo{pages}{124}.

\bibitem[{\citenamefont{{L. Grover}}({1996})}]{Grover1996}
\bibinfo{author}{\bibnamefont{{L. Grover}}}, in
  \emph{\bibinfo{booktitle}{{Proceedings of the Twenty-Eighth Annual Symposium
  on the Theory of Computing }}} (\bibinfo{publisher}{{ACM Press}},
  \bibinfo{address}{{New York}}, \bibinfo{year}{{1996}}), p.
  \bibinfo{pages}{212}.

\bibitem[{\citenamefont{{L. Grover}}(1997)}]{Grover1997a}
\bibinfo{author}{\bibnamefont{{L. Grover}}}, \bibinfo{journal}{Phys. Rev.
  Lett.} \textbf{\bibinfo{volume}{79}}, \bibinfo{pages}{325}
  (\bibinfo{year}{1997}).

\bibitem[{\citenamefont{{M. A. Nielsen and I. L. Chuang }}(2000)}]{Nielsen2000}
\bibinfo{author}{\bibnamefont{{M. A. Nielsen and I. L. Chuang }}},
  \emph{\bibinfo{title}{{Quantum computation and quantum information}}}
  (\bibinfo{publisher}{{Cambridge University Press}},
  \bibinfo{address}{{Cambridge}}, \bibinfo{year}{2000}).

\bibitem[{\citenamefont{{R. Jozsa and N. Linden}}(2003)}]{Jozsa2003}
\bibinfo{author}{\bibnamefont{{R. Jozsa and N. Linden}}},
  \bibinfo{journal}{Proc. R. Soc. London, Ser. A}
  \textbf{\bibinfo{volume}{459}}, \bibinfo{pages}{2011} (\bibinfo{year}{2003}).

\bibitem[{\citenamefont{{G. Vidal}}(2003)}]{Vidal2003}
\bibinfo{author}{\bibnamefont{{G. Vidal}}}, \bibinfo{journal}{Phys. Rev. Lett.}
  \textbf{\bibinfo{volume}{91}}, \bibinfo{pages}{147902}
  (\bibinfo{year}{2003}).

\bibitem[{\citenamefont{{D. Aharonov and M. Ben-Or}}(1996)}]{Aharonov1996}
\bibinfo{author}{\bibnamefont{{D. Aharonov and M. Ben-Or}}}, in
  \emph{\bibinfo{booktitle}{{Proceedings of the 37th Annual Symposium on the
  Foundations of Computer Science}}}, edited by
  \bibinfo{editor}{\bibnamefont{{S. Goldwasser}}} (\bibinfo{publisher}{{IEEE
  Computer Society}}, \bibinfo{address}{{Los Alamitos, CA}},
  \bibinfo{year}{1996}), p.~\bibinfo{pages}{46}.

\bibitem[{\citenamefont{{C. H. Bennett, H. J. Bernstein, S. Popescu, and B.
  Schumacher}}(1996)}]{Bennett1996a}
\bibinfo{author}{\bibnamefont{{C. H. Bennett, H. J. Bernstein, S. Popescu, and
  B. Schumacher}}}, \bibinfo{journal}{Phys. Rev. A}
  \textbf{\bibinfo{volume}{53}}, \bibinfo{pages}{2046} (\bibinfo{year}{1996}).

\bibitem[{\citenamefont{{C.H. Bennett, D.P DiVincenzo, J.A.Smolin and W.K
  Wootters}}(1996)}]{Bennett1996b}
\bibinfo{author}{\bibnamefont{{C.H. Bennett, D.P DiVincenzo, J.A.Smolin and W.K
  Wootters}}}, \bibinfo{journal}{Phys. Rev. A} \textbf{\bibinfo{volume}{54}},
  \bibinfo{pages}{3824} (\bibinfo{year}{1996}).

\bibitem[{\citenamefont{{S. Hill and W.K. Wootters}}(1997)}]{Hill1997}
\bibinfo{author}{\bibnamefont{{S. Hill and W.K. Wootters}}},
  \bibinfo{journal}{Phys. Rev. Lett.} \textbf{\bibinfo{volume}{78}},
  \bibinfo{pages}{5022} (\bibinfo{year}{1997}).

\bibitem[{\citenamefont{{W.K. Wootters}}(1998)}]{Wootters1998}
\bibinfo{author}{\bibnamefont{{W.K. Wootters}}}, \bibinfo{journal}{Phys. Rev.
  Lett.} \textbf{\bibinfo{volume}{80}}, \bibinfo{pages}{2245}
  (\bibinfo{year}{1998}).

\bibitem[{\citenamefont{{N. Linden, S. Popescu and W.K.
  Wooters}}(2002)}]{Linden2002a}
\bibinfo{author}{\bibnamefont{{N. Linden, S. Popescu and W.K. Wooters}}},
  \bibinfo{journal}{Phys. Rev. Lett.} \textbf{\bibinfo{volume}{89}},
  \bibinfo{pages}{207901} (\bibinfo{year}{2002}).

\bibitem[{\citenamefont{{N. Linden and W.K. Wooters}}(2002)}]{Linden2002b}
\bibinfo{author}{\bibnamefont{{N. Linden and W.K. Wooters}}},
  \bibinfo{journal}{Phys. Rev. Lett.} \textbf{\bibinfo{volume}{89}},
  \bibinfo{pages}{277906} (\bibinfo{year}{2002}).

\bibitem[{\citenamefont{{V. Vedral, M.B. Plenio, M.A. Rippin and P.L.
  Knight}}(1997)}]{Vedral1997}
\bibinfo{author}{\bibnamefont{{V. Vedral, M.B. Plenio, M.A. Rippin and P.L.
  Knight}}}, \bibinfo{journal}{Phys. Rev. Lett.} \textbf{\bibinfo{volume}{78}},
  \bibinfo{pages}{2275} (\bibinfo{year}{1997}).

\bibitem[{\citenamefont{{V. Vedral and M.B. Plenio}}(1998)}]{Vedral1998}
\bibinfo{author}{\bibnamefont{{V. Vedral and M.B. Plenio}}},
  \bibinfo{journal}{Phys. Rev. A} \textbf{\bibinfo{volume}{57}},
  \bibinfo{pages}{1619} (\bibinfo{year}{1998}).

\bibitem[{\citenamefont{{G. Vidal}}(2000)}]{Vidal2000}
\bibinfo{author}{\bibnamefont{{G. Vidal}}}, \bibinfo{journal}{{J. Mod. Opt}}
  \textbf{\bibinfo{volume}{47}}, \bibinfo{pages}{355} (\bibinfo{year}{2000}).

\bibitem[{\citenamefont{{M. Horodecki, P. Horodecki and R.
  Horodecki}}(2000)}]{Horodecki2000}
\bibinfo{author}{\bibnamefont{{M. Horodecki, P. Horodecki and R. Horodecki}}},
  \bibinfo{journal}{Phys. Rev. Lett.} \textbf{\bibinfo{volume}{84}},
  \bibinfo{pages}{2014} (\bibinfo{year}{2000}).

\bibitem[{\citenamefont{{V. Vedral, M. B. Plenio, K. Jacobs, and P. L.
  Knight}}(1997)}]{Vedral1997a}
\bibinfo{author}{\bibnamefont{{V. Vedral, M. B. Plenio, K. Jacobs, and P. L.
  Knight}}}, \bibinfo{journal}{Phys. Rev. A} \textbf{\bibinfo{volume}{56}},
  \bibinfo{pages}{4452} (\bibinfo{year}{1997}).

\bibitem[{\citenamefont{{H. Barnum and N. Linden}}(2001)}]{Barnum2001}
\bibinfo{author}{\bibnamefont{{H. Barnum and N. Linden}}}, \bibinfo{journal}{J.
  Phys. A} \textbf{\bibinfo{volume}{34}}, \bibinfo{pages}{6787}
  (\bibinfo{year}{2001}).

\bibitem[{\citenamefont{{M.S. Leifer, N. Linden and A.
  Winter}}(2004)}]{Leifer2004}
\bibinfo{author}{\bibnamefont{{M.S. Leifer, N. Linden and A. Winter}}},
  \bibinfo{journal}{Phys. Rev. A} \textbf{\bibinfo{volume}{69}},
  \bibinfo{pages}{052304} (\bibinfo{year}{2004}).

\bibitem[{\citenamefont{{O. Biham, M.A. Nielsen and T.
  Osborne}}(2002)}]{Biham2002}
\bibinfo{author}{\bibnamefont{{O. Biham, M.A. Nielsen and T. Osborne}}},
  \bibinfo{journal}{Phys. Rev. A} \textbf{\bibinfo{volume}{65}},
  \bibinfo{pages}{062312} (\bibinfo{year}{2002}).

\bibitem[{\citenamefont{{Y. Shimoni, D. Shapira and O.
  Biham}}(2004)}]{Shimoni2004}
\bibinfo{author}{\bibnamefont{{Y. Shimoni, D. Shapira and O. Biham}}},
  \bibinfo{journal}{Phys. Rev. A} \textbf{\bibinfo{volume}{69}},
  \bibinfo{pages}{062303} (\bibinfo{year}{2004}).

\bibitem[{\citenamefont{{Y. Shimoni, D. Shapira and O.
  Biham}}(2005)}]{Shimoni2005}
\bibinfo{author}{\bibnamefont{{Y. Shimoni, D. Shapira and O. Biham}}},
  \bibinfo{journal}{Phys. Rev. A} \textbf{\bibinfo{volume}{72}},
  \bibinfo{pages}{062308} (\bibinfo{year}{2005}).

\bibitem[{\citenamefont{{D. Shapira, Y. Shimoni and O.
  Biham}}(2006)}]{Shapira2006}
\bibinfo{author}{\bibnamefont{{D. Shapira, Y. Shimoni and O. Biham}}},
  \bibinfo{journal}{Phys. Rev. A} \textbf{\bibinfo{volume}{73}},
  \bibinfo{pages}{044301} (\bibinfo{year}{2006}).

\bibitem[{\citenamefont{{O. Biham, D. Shapira and Y.
  Shimoni}}(2003)}]{Biham2003}
\bibinfo{author}{\bibnamefont{{O. Biham, D. Shapira and Y. Shimoni}}},
  \bibinfo{journal}{Phys. Rev. A} \textbf{\bibinfo{volume}{68}},
  \bibinfo{pages}{{022326}} (\bibinfo{year}{2003}).

\bibitem[{\citenamefont{{A. Miyake and M. Wadati}}(2001)}]{Miyake2001}
\bibinfo{author}{\bibnamefont{{A. Miyake and M. Wadati}}},
  \bibinfo{journal}{Phys. Rev. A} \textbf{\bibinfo{volume}{64}},
  \bibinfo{pages}{042317} (\bibinfo{year}{2001}).

\bibitem[{\citenamefont{{W. D\"{u}r, G. Vidal and J.I. Cirac}}(2000)}]{Dur2000}
\bibinfo{author}{\bibnamefont{{W. D\"{u}r, G. Vidal and J.I. Cirac}}},
  \bibinfo{journal}{Phys. Rev. A} \textbf{\bibinfo{volume}{62}},
  \bibinfo{pages}{062314} (\bibinfo{year}{2000}).

\end{thebibliography}

\end{document}